\documentclass[prx,twocolumn,showpacs,floatfix,preprintnumbers,amsmath,amssymb,superscriptaddress]{revtex4}

\usepackage{graphicx}
\usepackage{amsmath}
\usepackage{amssymb}
\usepackage{color}
\usepackage{dcolumn}
\usepackage{epsfig}
\usepackage{bm}
\usepackage[urlcolor=blue]{hyperref}
\hypersetup{backref, colorlinks=true, linkcolor=blue, citecolor=blue}

\begin{document}
\title{Pressure tuning of superconductivity independent of disorder in Tl$_{2}$Ba$_{2}$CaCu$_{2}$O$_{8+\delta}$}

\author{Jian-Bo Zhang}
\affiliation{Department of Physics, South China University of Technology, Guangzhou 510640, China}
\affiliation{Center for High Pressure Science and Technology Advanced Research, Shanghai 201203, China}
\affiliation{Geophysical Laboratory, Carnegie Institution of Washington, Washington, DC 20015, U.S.A.}

\author{Viktor V. Struzhkin}
\affiliation{Geophysical Laboratory, Carnegie Institution of Washington, Washington, DC 20015, U.S.A.}

\author{Wenge Yang}
\affiliation{Center for High Pressure Science and Technology Advanced Research, Shanghai 201203, China}
\affiliation{Geophysical Laboratory, Carnegie Institution of Washington, Washington, DC 20015, U.S.A.}

\author{Ho-Kwang Mao}
\affiliation{Geophysical Laboratory, Carnegie Institution of Washington, Washington, DC 20015, U.S.A.}
\affiliation{Center for High Pressure Science and Technology Advanced Research, Shanghai 201203, China}

\author{Hai-Qing Lin}
\affiliation{Beijing Computational Science Research Center, Beijing 100089, China}

\author{Yong-Chang Ma}
\affiliation{School of Materials Science and Engineering, Tianjin University of Technology, Tianjin 300384, China}

\author{Nan-Lin Wang}
\affiliation{International Center for Quantum Materials, School of Physics, Peking University, Beijing 100871, China}
\affiliation{Collaborative Innovation Center of Quantum Matter, Beijing 100871, China}

\author{Xiao-Jia Chen}
\email{xjchen@hpstar.ac.cn}
\affiliation{Center for High Pressure Science and Technology Advanced Research, Shanghai 201203, China}
\affiliation{Geophysical Laboratory, Carnegie Institution of Washington, Washington, DC 20015, U.S.A.}

\date{\today}

\begin{abstract}
Varying the superconducting transition temperature over a large scale of a cuprate superconductor is a necessary step for identifying the unsettled mechanism of superconductivity. Chemical doping or element substitution has been proven to be effective but also brings about lattice disorder. Such disorder can completely destroy superconductivity even at a fixed doping level. Pressure has been thought to be the most clean method for tuning superconductivity. However, pressure-induced increase of disorder was recognized from recent experiments. By choosing a disordered Tl$_{2}$Ba$_{2}$CaCu$_{2}$O$_{8+\delta}$ at the optimal doping, we perform single-crystal x-ray diffraction and magnetic susceptibility measurements at high pressures. The obtained structural data provides evidence for the robust feature for the disorder of this material in the pressure range studied. This feature ensures the pressure effects on superconductivity distinguishable from the disorder. The derived parabolic-like behavior of the transition temperature with pressure up to near 30 GPa, having a maximum around 7 GPa, offers a platform for testing any realistic theoretical models in a nearly constant disorder environment. Such a behavior can be understood when considering the carrier concentration and the pairing interaction strength as two pressure intrinsic variables. 
\end{abstract}

\maketitle

There has been no agreement on what mechanism controls superconductivity in the cuprate superconductors since their discovery. Superconductivity in cuprates usually appears at certain doping levels in which spin and charge stripes \cite{tran}, pseudogap \cite{ding,yhe,kfuj}, fluctuation of superconductivity \cite{zaxu}, spin glass \cite{kohs}, charge order \cite{twu,ghir,chan}, and other orders \cite{lake,torc} coexist and/or compete with superconductivity. Whether a general principle underlies the relationship between superconductivity and these competing orders is an outstanding question. Finding its answer(s) would eventually revolutionize condensed matter physics. Meanwhile, determining the nature of the competing order can provide information about the mechanism of pairing in the superconducting state. Even an empirical connection between superconductivity and competing orders could offer new opportunities to the search for novel superconducting materials.

Amongst various known orders, lattice disorder has been generally observed to significantly affect the superconducting transition temperature $T_{c}$ of many cuprates \cite{att,16,12,13,19,eisa,mura,17,20}. Disorder is usually introduced into the cation sites in the plane adjacent to the CuO$_{2}$ plane through the cation substitution \cite{12,13,19,eisa,mura,17,20} and even the change of the size variance \cite{att,16} with the hole density kept constant. In the La$_{2}$CuO$_{4}$-based system, synchrotron x-ray and neutron-diffraction measurements demonstrated that the cation disorder induces low-temperature structural instabilities which reorient the tilting of the copper-oxygen octahedra \cite{12}. At a fixed doping level, $T_{c}$ was found to have the highest value for flat and square CuO$_{2}$ planes in the tetragonal structure but it is reduced by the structural distortions of the CuO$_{2}$ planes in the orthorhombic one \cite{13}. Fixing the mean A-site cation radius of the optimally doped compounds, the only change of the size variance also can rapidly suppress $T_{c}$ in the same manner as the substitution of diamagnetic elements such as Zn for Cu \cite{att,16}. This new kind of disorder was later suggested to work as weak scatterers in contrast to the substitution of Cu by Zn \cite{19}. The disorder-tuned bulking of CuO$_{2}$ was thus believed to account for the $T_{c}$ variation in La$_{2}$CuO$_{4}$-based system \cite{12,13}. The single-layer Bi$_{2}$Sr$_{2}$CuO$_{6+\delta}$ was suggested to share the similar mechanism for the disorder effect on superconductivity \cite{19}. However, a systematic study \cite{eisa} on both Bi$_{2}$Sr$_{2}$CuO$_{6+\delta}$ and Bi$_{2}$Sr$_{2}$CaCu$_{2}$O$_{8+\delta}$ revealed that the cation disorder is in reality associated with the chemical inhomogeneities. The superconducting gap of the latter was reported to be suppressed with increasing disorder mainly in the nodal region while the antinodal gap remains almost unchanged \cite{mura}. The same phase separation was also proposed as the origin of the out-of-plane disorder on superconductivity in YBa$_{2}$Cu$_{3}$O$_{7-\delta}$ (Y123)-based materials \cite{17,20}. Therefore, the impact of disorder on the electronic structure and superconducting properties in high-$T_{c}$ cuprates is still not clear, a task which is very difficult even in conventional superconductors \cite{beli}.

Changing lattice disorder was originally thought to provide direct information on the pairing interaction strength for high-$T_{c}$ cuprate \cite{12,13} because the disorder is independent on the carrier density in these systems. The observation of disorder-induced chemical inhomogeneities or phase separation \cite{eisa,17,20} adds the complexity of the effects of disorder on superconductivity. Searching for a cleaner variable to tune and control superconductivity at fixed doping level is emergent and attractive. Pressure was recognized to be best variable for satisfying all these demands. The developments of high-pressure techniques in the past half a century has already enabled the measurements at ambient conditions available at high pressures. In fact, superconductivity studies \cite{30,Struzhkin,Eremets,Chen} benefited a lot from these technique developments. The record high $T_{c}$ of 164 K in cuprates \cite{lgao} and the new record for superconductivity at 190 K in H-S system \cite{droz} were achieved at high pressures. Interestingly, Calamiotou {\it et al}. \cite{Calamiotou,Gantis} emphasized the disorder effects on superconductivity under pressure in two superconductors YBa$_{2}$Cu$_{4}$O$_{8}$ (Y124) and Y123 and nonsuperconducting PrBa$_{2}$Cu$_{3}$O$_{6.92}$ by combining synchrotron x-ray powder diffraction and Raman spectroscopy measurements. Their analysis strongly shows a clear anomaly in the evolutions of both the lattice parameters and phonon modes with pressure in these systems. The disorder was observed to start increasing with pressure when the anomaly appears. The observed phase separation is analogous to the one previously reported by the same authors in these systems at ambient pressure \cite{17,20}. Recently, Nakayama {\it et al}. \cite{japan} investigated the crystal structure and electrical resistivity of Y124 under pressure up to 18 GPa. A dramatic change of $T_{c}$ was observed to be accompanied by a structural phase transition around 10 GPa. This study is followed by Raman spectroscopy and $ab~initio$ calculations \cite{souliou}, which confirm the phase transition. Pressure-induced phase separation or phase transition calls for a careful examination of the roles of pressure, disorder, and phase separation/transition on superconductivity. The issue regarding whether the enhanced disorder could happen by the application of pressure and whether the phase separation/transition at high pressures is unique in other cuprates becomes crucial for understanding the mechanism of superconductivity in high-$T_{c}$ cuprates.

Lattice instability and structural fluctuation have been generally observed in thallium-based cuprate superconductors \cite{Koyama,Zetterer,10,Y} through the measurements of electron-diffraction, far-infrared reflectivity, and Cu K-edge extended x-ray-absorption fine structure, suggesting a close connection between lattice degrees of freedom and superconductivity. This superconducting family thus offers a good opportunity to examine the lattice disorder effects on superconductivity. In this work, we choose a well characterized bilayer Tl$_{2}$Ba$_{2}$CaCu$_{2}$O$_{8+\delta}$ (Tl2212) single crystal and investigate the pressure effects on lattice evolution and superconductivity to address the above mentioned issues. Our central experimental finding is that the disorder of this superconductor does not develop with pressure from its initial state at ambient pressure, though $T_{c}$ undertakes a huge change over the pressure range studied. Our results indicate that the pressure dependence on superconductivity is controlled by the intrinsic parameters themselves and is barely affected by the disorder in the studied system.

The optimally doped Tl2212 single crystal with $T_{c}$ of 109\,K was grown by the flux method \cite{29}. A piece of high-quality crystal together with a small ruby ball is loaded to a symmetric diamond anvil cell with culet of 300 $\mu$m for the single crystal x-ray diffraction (XRD) measurement under pressure. The sample chamber of about 110 $\mu$m in diameter was created in a gasket made by the stainless steel. Neon gas was loaded into the sample chamber as the pressure medium using the GSECARS gas loading system at the Advanced Photon Source, Argonne National Laboratory \cite{32}. Good hydrostaticity was maintained to the highest pressure measured and pressure was gauged by fluorescence line of ruby at room temperature \cite{31}. High-quality single crystal and good hydrostaticity environment ensure the collection of the structural data in high quality.

\begin{figure}[tbp]
\includegraphics[width=\columnwidth]{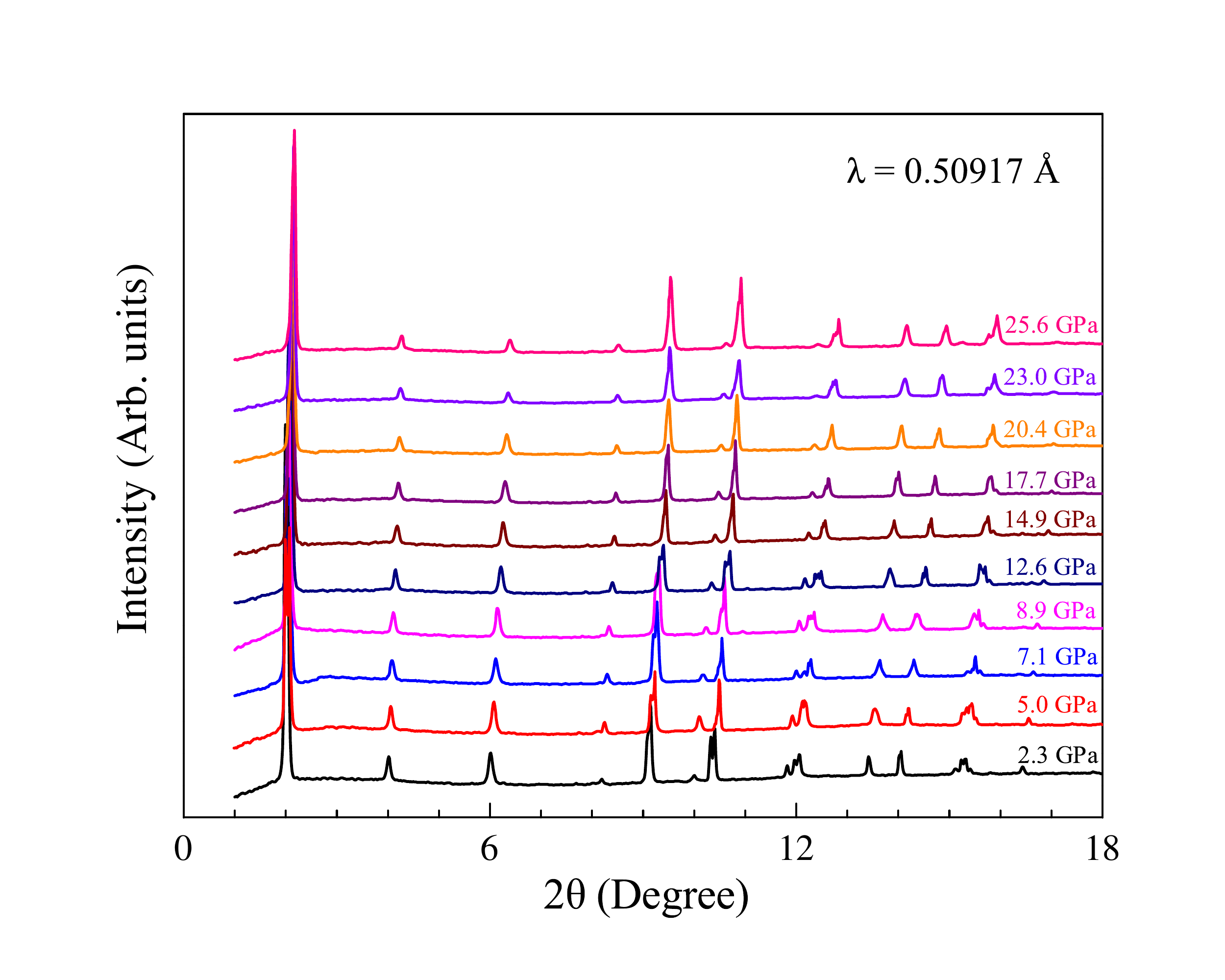}
\caption{The integrated single crystal x-ray diffraction data of the optimally doped Tl$_{2}$Ba$_{2}$CaCu$_{2}$O$_{8+\delta}$ at various pressures up to 25.6 GPa.}
\end{figure}

We first investigate the evolution of disorder with pressure by carefully analyzing the crystal structure and lattice parameters. Pressure-dependent single crystal XRD patterns were collected by a MAR345 image plate at pressures up to $25.6$ GPa at room temperature, by using Beamline at 16BM-D of High Pressure Collaborative Access Team at the Advanced Photon Source, with an incident x-ray wavelength of 0.50917 \AA. The two dimensional patterns were integrated into one dimensional XRD patterns with Fit2D software \cite{33}. The single crystal XRD data for the optimally doped Tl2212 at various pressures and room temperature is shown in Fig. 1. There are no new diffraction peaks and no peak merging and/or splitting. The peak intensity and their full width at half maximum (HWHM) are insensitive to the applied pressure. The diffraction peaks monotonically shift to higher angles slowly with increasing pressure, illustrating the homogeneous compression of the lattice. All these behaviors indicate that the structure of Tl2212 was stable under pressure. This result is different to the pressure-induced phase separation observed in Y123 superconductor \cite{Gantis}, in which new diffraction peaks appear at pressure around 3.7 GPa and both the widths and frequencies of some selected vibrational modes show abnormal behaviors at pressure range of 3.7 and 10 GPa. Our diffraction data does not support pressure-induced phase transition or phase separation in Tl2212.

The structure of Tl2212 has been refined from the single crystal XRD and neutron powder diffraction measurements \cite{38,39}. Its unit cell is tetragonal with a space group of $I4/mmm$. The inset of Fig. 2 presents the structure of Tl2212. As can be seen, Tl2212 consists of two insulating Tl-O and Ba-O blocking layers and two structurally equivalent Cu-O plans seperated by Ca layer. The lattice parameters are $a$ = 3.85 \AA~and $c$ = 29.3 \AA~at ambient pressure. Our diffraction patterns collected at high pressures do not exhibit substantial change compared to those collected at ambient pressure. The intensity-$vs$-2$\theta$ diffraction patterns were analyzed in terms of the Le Bail method by using GSAS software \cite{34,Toby}. All patterns were fitted well with the tetragonal structure with the space group of $I4/mmm$ up to $\sim 25.6$ GPa. Figure 2 illustrates a typical Le Bail fitting of the single crystal XRD data of the optimally doped Tl2212 at pressure of 2.3 GPa and refinement factors for this fitting are $R_{p}=1.4 \%$, $R_{wp}=2.8 \%$, and $\chi^{2} = 4.8 \%$, within the error bar of the experiment. We also fitted the higher pressure date with the same $I4/mmm$ and the results show that the diffraction patterns could be clearly indexed with this structure. These results indicate that Tl2212 remains in the tetragonal structure at the pressure regime studied.

\begin{figure}[tbp]
\includegraphics[width=\columnwidth]{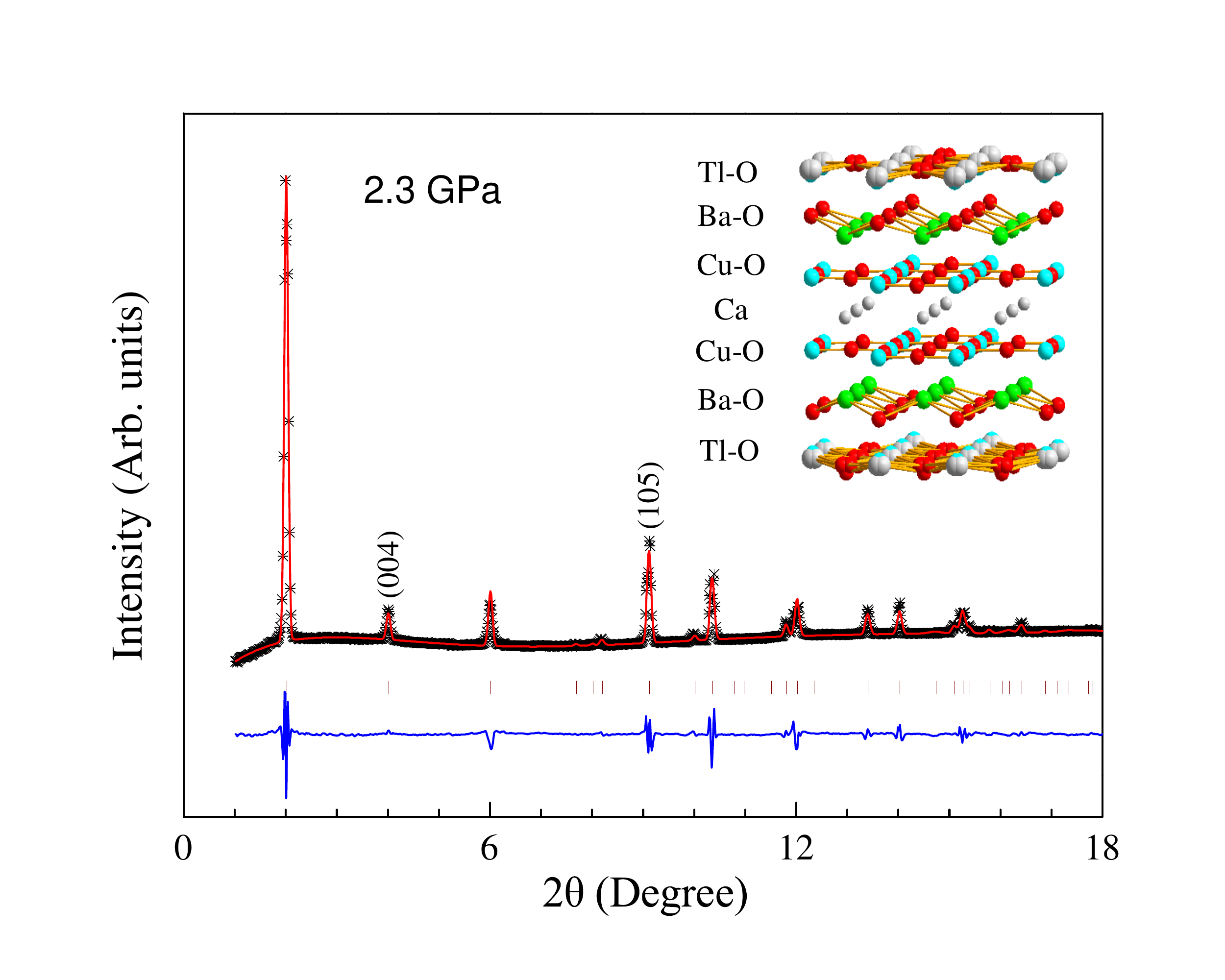}
\caption{The observed single crystal x-ray diffraction pattern (stars), Le Bail fit (upper continuous red line), and difference between the observed and calculated profiles (bottom blue line) obtained after Le Bail fitting of the optimally doped Tl$_{2}$Ba$_{2}$CaCu$_{2}$O$_{8+\delta}$ at 2.3 GPa based on the tetragonal structure with the space group of $I4/mmm$. The middle sticks refer to the peak positions. (004) and (105) are two characteristic peaks. Inset: Schematic of the crystal structure.}
\end{figure}

The evolution of disorder with pressure was usually studied by combining XRD and Raman measurements \cite{Calamiotou,Gantis}. The change in disorder with pressure is often judged by three basic facts: (i) The increase of microstrains with the application of pressure; (ii) Pressure-induced lattice anomalies (the lattice parameters and cell volume exhibit obvious deviations from the expected equation of state); (iii) The unconventional variation of both the phonon frequencies and widths of some modes with increasing pressure. The existence of the disorder in the thallium-based cuprates at ambient pressure has been reported previously \cite{Koyama,Zetterer,10,Y}. However, it is unclear how pressure would affect the disorder in these compounds. Here we examine the disorder effects on Tl2212 based on these three aspects.

We chose two characteristic peaks of (004) and (105) to represent the $c$ axis and all three axes, respectively. We calculated the evolution of the microstrain of these peaks with pressure. When evaluating the degree of microstrain disorder, we fitted the two diffraction peaks by using the pseudo-Voigt function, assuming a linear combination of Gaussian and Lorentzian components. We thus obtained the integral breadth $\beta$. The microstrain was then calculated through $\varepsilon = \beta/(4\tan\theta)$, where $\theta$ is the diffraction angle. In this way, the absolute values of $\varepsilon$ may contain a contribution from the instrument, but this effect is independent on pressure and does not influence the evolution of disorder with pressure. The FWHMs for peaks (004) and (105) are insensitive with increasing pressure. The calculation results indicate that the microstrain disorder does not enhance with increasing pressure (Fig. 3). We also chose other peaks to evaluate the microstrain and obtained the similar results. This situation is different to the abnormal enhancement of the microstrain together with the broadening FWHMs of diffraction peaks in superconducting Y124 and Y123 above 3.7 GPa \cite{Calamiotou,Gantis}. Our results indicate that pressure has negligible effect on the disorder in the optimally doped Tl2212 in the pressure range studied.

\begin{figure}[tbp]
\includegraphics[width=\columnwidth]{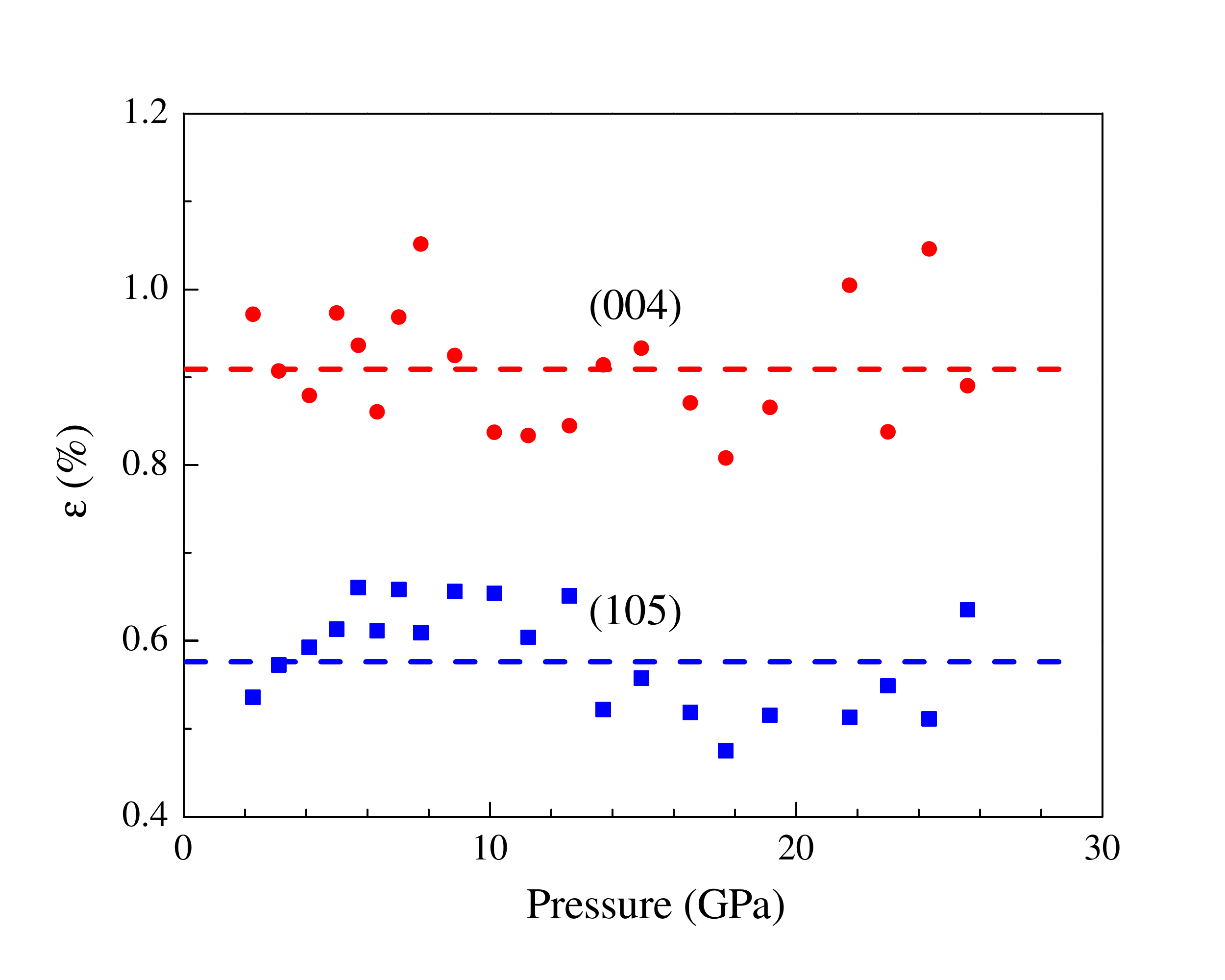}
\caption{Pressure dependence of the microstrain disorder for the two major diffraction peaks (004) and (105) in the optimally doped Tl$_{2}$Ba$_{2}$CaCu$_{2}$O$_{8+\delta}$. The dashed lines are the guide to eyes.}
\end{figure}

Figure 4 presents the pressure dependence of the lattice parameters along the $a$ and $c$ axes and pressure-volume ($P$-$V$) equation of state, respectively. As can be seen, all the lattice parameters exhibit continuous compressibility with increasing pressure, contributing to a monotonous decrease of the unit-cell volume upon compression. The reductions as large as 4.4\% and 6.8\% are observed along the $a$ and $c$ axes at 25.6 GPa, respectively [Fig. 4(a)]. The similar anisotropic compressibility has also been observed in Tl2212 and Tl$_{2}$Ba$_{2}$Ca$_{2}$Cu$_{3}$O$_{10+\delta}$ \cite{42,fiez}. The line in Fig. 4(b) is the fitting result of the phase using the third-order Birch-Murnaghan equation of state \cite{41}:
\begin{equation}
P=3B_0f_E(1+2f_E)^{\frac{5}{2}}[1+\frac{3}{2}(B_0^{'}-4)f_E],
\end{equation}
where $f_E=[(\frac{V_0}{V})^{\frac{2}{3}}-1]$, $V_0$ is the volume per formula unit at ambient pressure, $V$ is the volume at pressure $P$ given in GPa, $B_0$ is the bulk modulus at ambient pressure, and ${B_0'}$ is value of the first derivative of bulk modulus with respective to pressure. In contrast to Y124 and Y123  \cite{Calamiotou,Gantis}, the measured cell volume of Tl2212 does not obviously deviate from the expected equation of state and the compression of the lattice parameters is smoothly developed with pressure. Our high-quality structural data again does not indicate the pressure-induced increase of the disorder in Tl2212.

A least-square fit to the measured $P$-$V$ data yields $B_0$ = 111.7 $\pm$ 5.0 GPa for Tl2212 with a fixed $B_0'$ = 4.0, while $V_{0}$ = 438.8 $\pm$ 0.8 {\AA}$^{3}$. A bulk modulus of 233 GPa was reported for Tl2212 from early energy dispersive XRD measurements \cite{42}. This value is almost twice of ours for Tl2212. However, our obtained result is comparable to the reported 137 GPa for the trilayer Tl$_{2}$Ba$_{2}$Ca$_{2}$Cu$_{3}$O$_{10+\delta}$ \cite{fiez}. The smaller bulk moduli of 73, 62.5, and 68.56 GPa were also reported for sister bilayer Bi-based system \cite{taji,stau,jdyu}. Our measurements were conducted by the single crystal XRD technique with neon as the pressure transmitting medium. The obtained results should capture the essential structure information of the studied compound.

Raman spectroscopy is another effective method for investigating the phase separation/transition and lattice distortions. The deviation of certain vibrational modes from the expected monotonic evolution with pressure or the sudden change of the peak width are often taken as evidence for the change of the disorder \cite{d3,d5}. For the studied Tl2212, the systematic and monotonic variation of all oxygen A$_{1g}$ phonons was reported with increasing pressure up to 25 GPa from two independent Raman measurements \cite{40,v}. There is no indication of peak splitting or merging at higher pressures. Thus, the possibility for phase separation or phase transition in this material at the pressure range studied can be ruled out.  

\begin{figure}[tbp]
\includegraphics[width=\columnwidth]{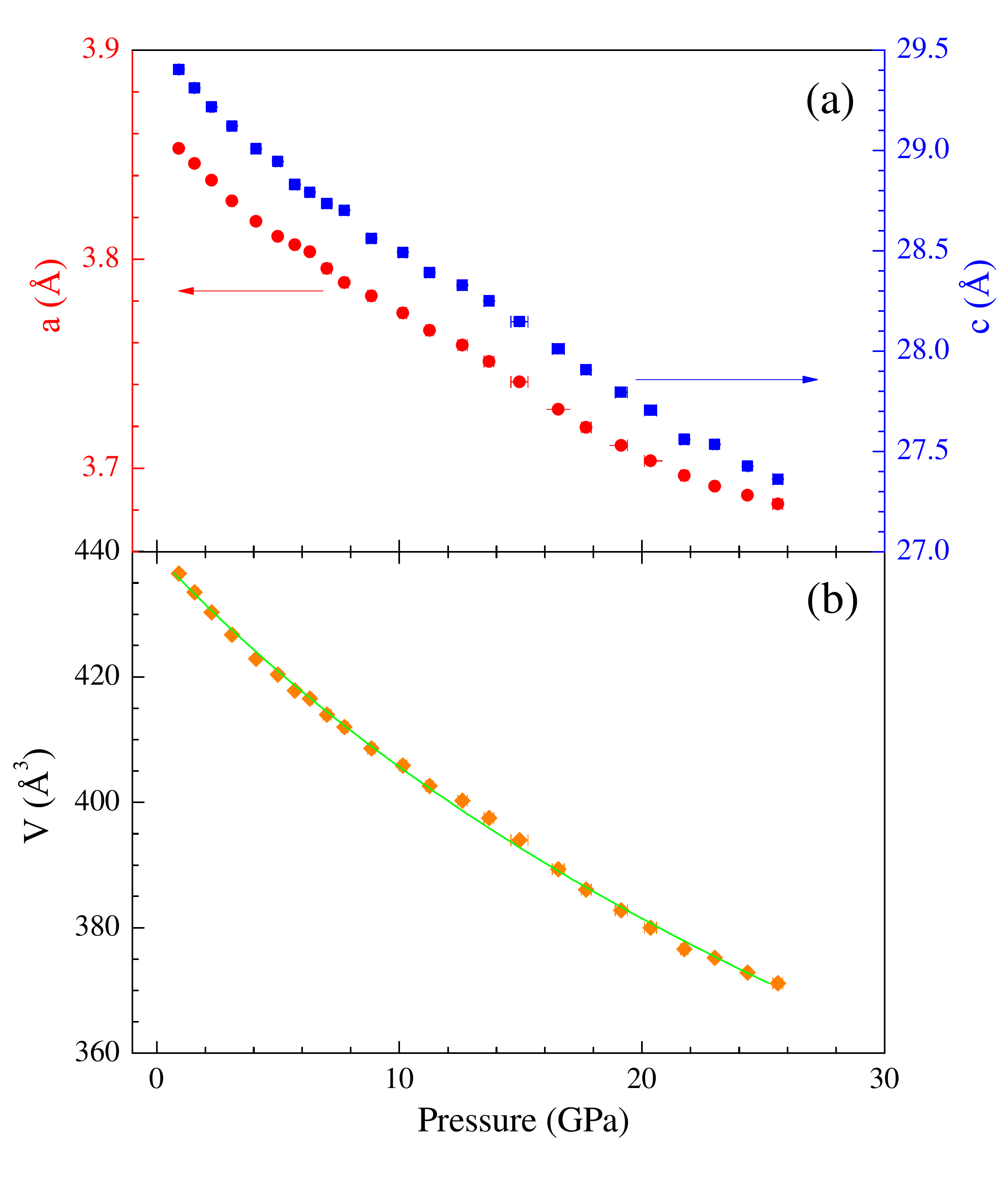}
\caption{The lattice parameters along the $a$ and $c$ axes (a) and the unit-cell volume (b) of the optimally doped Tl$_{2}$Ba$_{2}$CaCu$_{2}$O$_{8+\delta}$ as a function of pressure up to 25.6 GPa. The solid points are the measured data and the curve is the fit of the $P$-$V$ equation of state to the data.}
\end{figure}

The analysis of both the high-pressure structural and Raman data has established that disorder is insensitive to pressure in Tl2212. There does not exist phase transition or phase separation in Tl2212 in the pressure range studied. Therefore, the concern of pressure-induced effect of disorder in the studied Tl2212 can be safely removed. The optimally doped Tl2212 offers an excellent opportunity for examining the intrinsic pressure effects on superconductivity. The consistent experimental data based on the same single crystal can provide a testing platform for any existing theoretical models. Next we present the results from a systematic high-pressure study on $T_{c}$ for Tl2212 by using the same crystal at the optimal doping as the structural study.

For the $T_{c}$ measurements, a single crystal with dimensions of $120\times100\times20$ $\mu$$m^{3}$ was loaded into a Mao-Bell cell which was made from hardened Be-Cu alloy. A nonmagnetic Ni-Cr alloy gasket was preindented to 35 $\mu$m thick with a hole of 250 $\mu$m in diameter to serve as the sample chamber. Daphne 7373 was loaded into the gasket hole as the pressure transmitting medium. We used a highly sensitive magnetic susceptibility technique to determine $T_{c}$ at high pressures, as detailed previously \cite{30,Struzhkin,Eremets,Chen}. Pressure was gauged by the shift of the R1 fluorescence line of ruby \cite{31}.

\begin{figure}[tbp]
\includegraphics[width=\columnwidth]{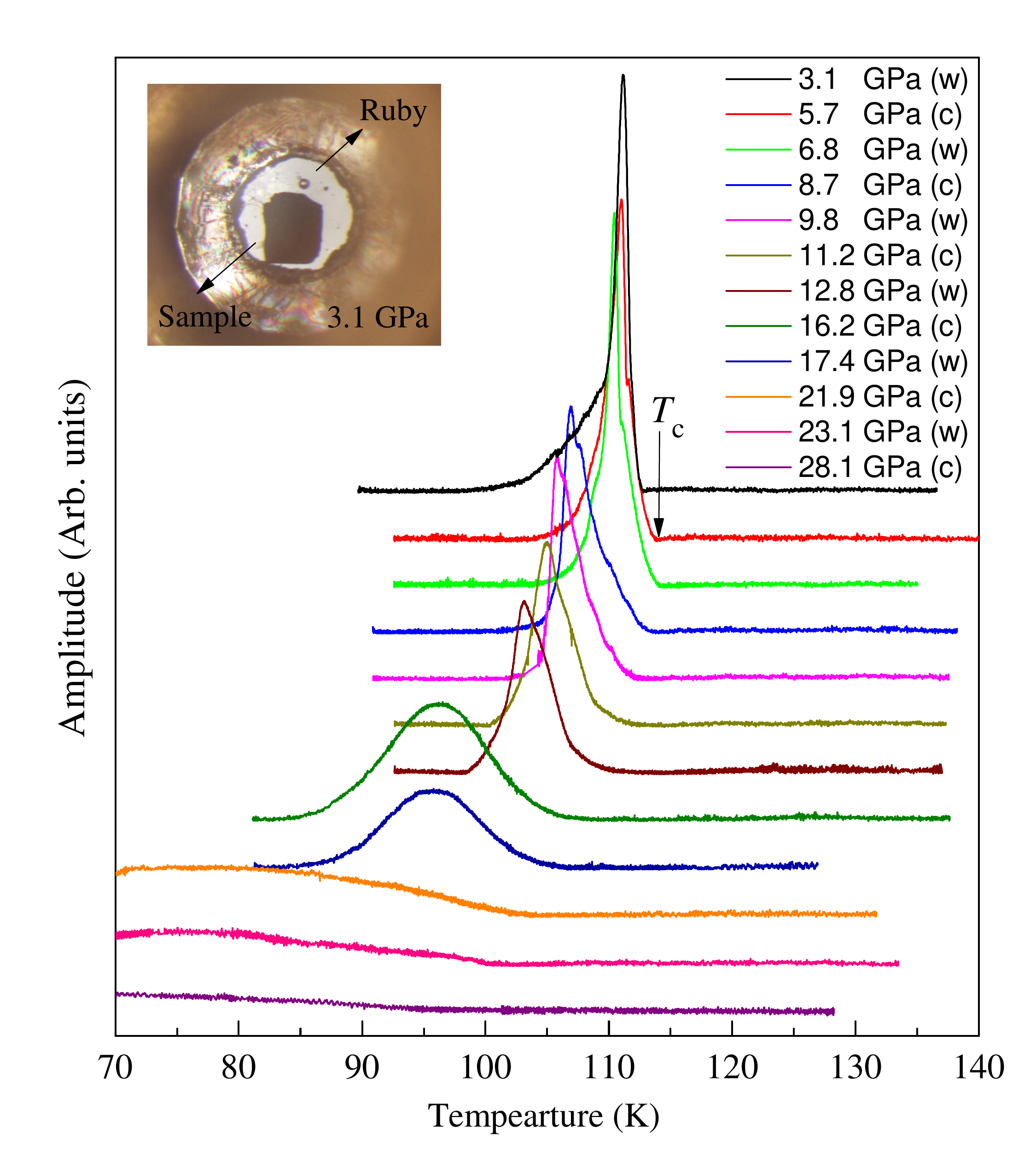}
\caption{Magnetic susceptibility signal for the optimally doped Tl$_{2}$Ba$_{2}$CaCu$_{2}$O$_{8+\delta}$ single crystal measured at various pressures up to 28.1 GPa. Inset: The sample and ruby in Daphne 7373 environment in the gasket hole at pressure of 3.1 GPa.}
\end{figure}

A photograph taken through the diamond windows of a single crystal at 3.1 GPa is shown in the inset of Fig. 5. The crystal, together with a small ruby ball, was put in a Daphne 7373 environment in the gasket hole. Large sample was helpful for getting strong amplitude signal. Daphne 7373 offers hydrostaticity and protection for the sample to keep the bulk feature during measurements at high pressures. Figure 5 shows the representative temperature scans at different applied pressures. For each pressure run, the signal was measured during both cooling and warming cycles at low temperature below 140\,K. The pressure was applied and measured at low temperature. Superconducting transition is identified as the temperature where the signal begins to develop on the high-temperature side \cite{Chen}. The superconducting transition of 109\,K was obtained at ambient pressure. It is clear that at the pressure of 6.8 GPa the superconducting transition shifts to higher temperature (114 K), but it returns beyond that pressure. The similar weakening of the amplitude and broadening of the width of the signal upon heavy compression have also been observed in other cuprate superconductors \cite{53,wang}.

Figure 6 summarizes the evolution of $T_{c}$ of the optimally doped Tl2212 with pressure. $T_{c}$ is initially increased with applied pressure and reaches a maximum of $\sim$ 114\,K around 6.8 GPa, then decreases at higher pressures. Although the amplitude of the single crystal becomes weak with the applied pressure, we still can distinguish the sample signal crystal from the background at pressures up to 28 GPa. The similar parabolic-like behavior with a maximum $T_{c}$ around 2.5 GPa (2.0 GPa) has been observed from early experiments by using $ac$ magnetic susceptibility \cite{all} (resistance \cite{35}) technique. There is only modest increase of $T_{c}$ from the initial value to the maximum for both the experiments with the highest pressure less than 4 and 8 GPa, respectively. The initial pressure derivative of $T_{c}$ (${dT_{c}/dP}$) is about 0.75 KGPa$^{-1}$ from our measurements. This value is smaller compared to the early measurements at relatively low pressures \cite{all,35,36,37}. These differences might origin from the samples with different dopants, sample qualities, or pressure media. The critical pressure for the occurrence of the maximum $T_{c}$ and parabolic-like behavior are similar to those in the sister optimally doped Bi$_{2}$Sr$_{2}$CaCu$_{2}$O$_{8+\delta}$ by using the same measurement technique \cite{53}. Combining the structural information, we established the consistent evolution of the structure and $T_{c}$ with pressure near 30 GPa on the same crystal of the optimally doped Tl2212. The experimental data on high-quality sample enables the potential test of any realistic theoretical models.

\begin{figure}[tbp]
\includegraphics[width=\columnwidth]{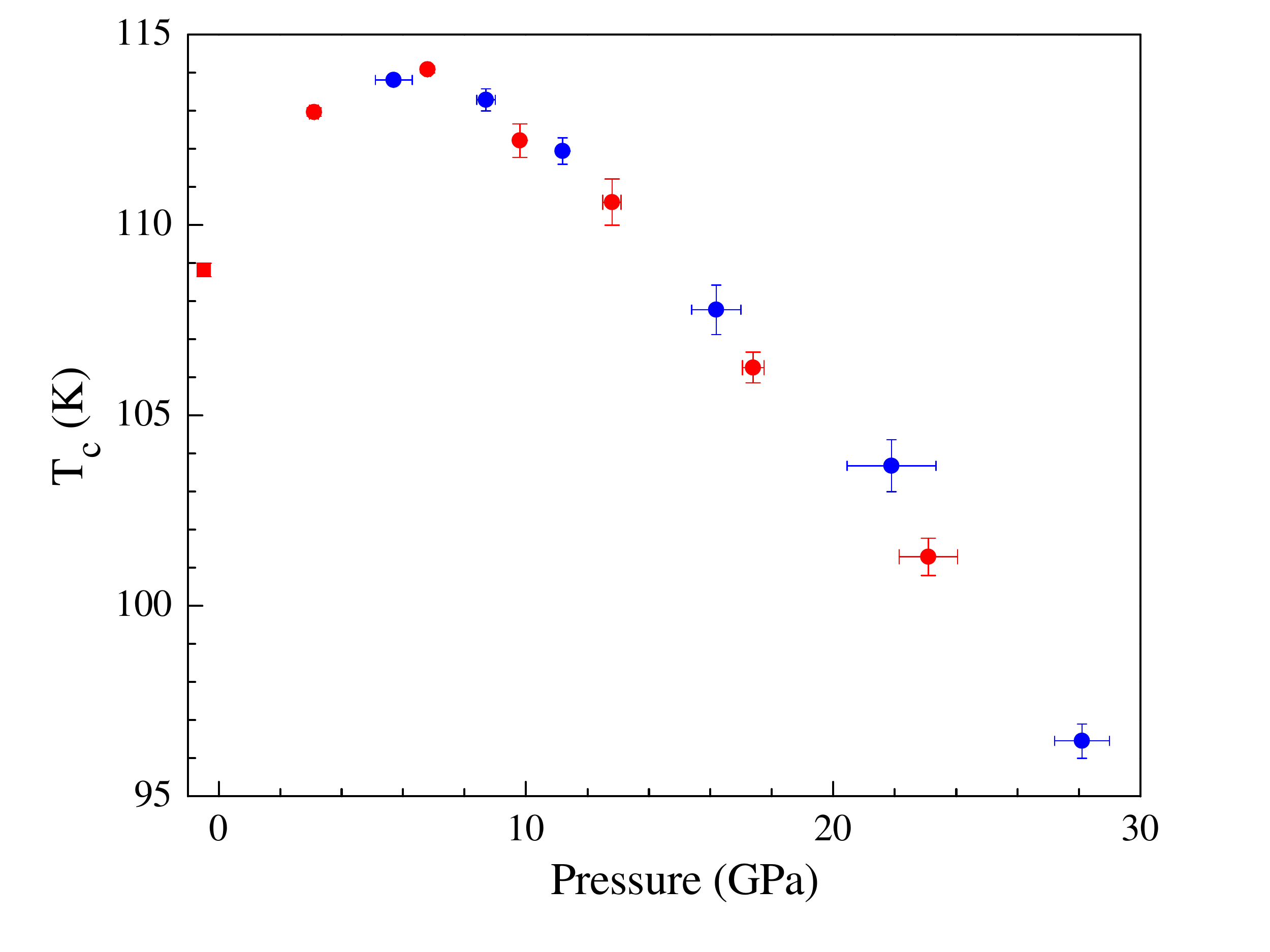}
\caption {Pressure dependence of $T_{c}$ for the optimally doped Tl$_{2}$Ba$_{2}$CaCu$_{2}$O$_{8+\delta}$ single crystal. The red (blue) solid circles represent the measurements in the warming (cooling) cycle, respectively. Red square represents $T_{c}$ at ambient pressure.}
\end{figure}

The obtained parabolic-like $T_{c}$ vs $P$ behavior for the optimally doped Tl2212 with 5 K enhancement of $T_{c}$ at a critical pressure about 7 GPa is generic for almost all optimally doped cuprates. This behavior can be explained by many two-component models including charge carrier concentration and another intrinsic variable. For the generally used charge-transfer model, the maximum of $T_{c}$ is considered as the other intrinsic variable besides carrier concentration. This simple model worked very well for many cuprates with different dopants \cite{cjiao,cgong}. The developed models based on BCS-like gap equations also can be used to reproduce the $T_{c}$ variation with pressure for many compounds \cite{26,27}. The later developments for identifying the pairing interaction strength as the second intrinsic variable has led to many interesting explanations for the $T_{c}$ evolution with pressure and its pressure derivatives at different dopants \cite{28,53,cyu}, the uniaxial pressure effect on $T_{c}$ \cite{cyin}, the rare-earth ionic size effect on $T_{c}$ \cite{cyu,csu}, and even the strain effect on $T_{c}$ \cite{clg}. This variable was recently confirmed by high-pressure NMR measurements \cite{52}. Currently, lattice vibrations and excitations of electronic origin such as spin or electric polarizability fluctuations are generally considered potential candidates of Cooper pairing in the cuprate superconductors. Both were found to have important contributions to the pairing interaction strengths \cite{dal,chia}. We have demonstrated that the consideration of pairing interaction strength and carrier concentration is also sufficient for explaining the observed $T_{c}$ behavior at high pressures in cuprates even within the framework of phonon-mediated pairing \cite{cpnas,cprb}. The similar two components from the dynamic inhomogeneity-induced pairing model \cite{43} including the pairing scale, which characterizes pair formation and its proportional to the energy gap, and the phase ordering scale, which controls the stiffness of the system to phase fluctuations and is determined by the superfluid density, have also been used to explain the experiments \cite{Chen}. In fact, the superfluid density is approximately proportional to the carrier concentration before the optimal level and the pairing interaction strength should be scaled by the energy gap. Therefore, both the carrier concentration and the pairing interaction strength are two well determined intrinsic pressure variables at least for cuprate superconductors.   

Now we would like to highlight these two intrinsic pressure variables for the studied Tl2212 and other cuprates. The pressure-induced reduction of the Hall coefficient has been observed for many cuprates \cite{48}, indicating the increase of the carrier  concentration in the CuO$_{2}$ plane. The charge-transfer process can also be monitored by investigating the vibrational frequency of the apical oxygen \cite{46}. The vibration frequency of the apical oxygen in Tl2212 was observed to get hardening with increasing pressure \cite{40,v}, resulting in the charge transfer from the charge reservoir to the CuO$_{2}$ plane. This behavior has been supported by the high-pressure Hall coefficient measurements on Tl-based cuprates \cite{48}. 

Recent experimental efforts shed new light on the nature of the pairing interactions \cite{mere,mans}. The typical scale of 2.6 eV was attributed to a fingerprint of ``Mottness'' in the superconducting state. This energy scale, set by the superexchange interaction, was found to control all $T_{c}$'s of a cuprate system \cite{mere}. If the superexchange interaction indeed provides the driving force for superconductivity in cuprates, one can readily learn the pressure effect on the pairing interaction. For almost all cuprate families, the superexchange interaction was clearly found to increase with increasing pressure from Raman scattering measurements \cite{49,50,51}. The pairing interaction strength as another intrinsic variable is again supported from these experiments. Our observed parabolic-like $T_{c}$ vs $P$ behavior for the optimally doped Tl2212 is thus the result of the interplay between the carrier concentration and the pairing interaction strength.

In summary, the consistent information on the evolution of both the structure and superconducting transition temperature with pressure up to near 30 GPa has been obtained for the optimally doped Tl$_{2}$Ba$_{2}$CaCu$_{2}$O$_{8+\delta}$ by using a high-quality single crystal. The structural data collected from single-crystal x-ray diffraction measurements provides evidence for the robust feature for the disorder of this material under pressure. This allows the investigation of the pressure effects on superconductivity distinguishable from the disorder. Highly sensitive magnetic susceptibility measurements on the same crystal as used for structural study yielded a generic parabolic-like $T_{c}$ behavior upon compression with a maximum around 7 GPa. We demonstrated that this behavior can be explained by considering  the carrier concentration and the pairing interaction strength as two pressure intrinsic variables. The optimally doped Tl$_{2}$Ba$_{2}$CaCu$_{2}$O$_{8+\delta}$ was found to still hold 89\% of its ambient value of $T_{c}$ when reducing 15\% of the unit-cell volume near 30 GPa in a nearly constant disorder environment. This advantage will be helpful for any realistic test for the $T_{c}$ variation of theoretical models in a clean environment. 

This work was supported by EFree, an Energy Frontier Research Center funded by the U.S. Department of Energy (DOE), Office of Science, Office of Basic Energy Sciences (BES) under Award Number DE-SG0001057. The XRD measurements were supported by the Cultivation Fund of the Key Scientific and Technical Innovation Project Ministry of Education of China (Project No. 708070). The magnetic susceptibility measurements were supported by the DOE under Grant No. DE-FG02-02ER45955. The sample design and growth were supported by the Natural Science Foundation of China (Nos. 11120101003 and 11327806). The theoretical analysis was supported by NSAF (No. U1230202). The use of HPCAT, APS was supported by Carnegie Institute of Washington, Carnegie DOE Alliance Center, University of Nevada at Las Vegas, and Lawrence Livermore National Laboratory through funding from DOE-National Nuclear Security Administration, DOE-Basic Energy Sciences, and NSF. APS is supported by DOE-BES, under Contract No. DE-AC02-06CH11357.

\end{document}